\documentstyle[12pt,fleqn]{article}
\topmargin - 0.8cm

\renewcommand{\theequation}{\arabic{section}.\arabic{equation}}
\renewcommand{\thesection}{\arabic{section}.}

\catcode`@=11
\@addtoreset{equation}{section}
\catcode`@=12
\mathchardef\SGamma="7100

\begin{document}
\title{\vskip-1.7cm \bf  Geometry of the Dirac and Reduced Phase Space
Quantization of Systems Subject to First Class Constraints}
\date{}
\author{A.O.Barvinsky}
\maketitle
\hspace{-8mm}{\em
Theory Department, Lebedev Physics Institute and Lebedev Research Center in
Physics, Leninsky Prospect 53,
Moscow 117924, Russia}
\begin{abstract}
Geometric properties of operators of quantum Dirac constraints and
physical observables are studied in semiclassical theory of generic
constrained systems. The invariance transformations
of the classical theory -- contact canonical transformations and arbitrary
changes of constraint basis -- are promoted to the quantum domain as
unitary equivalence transformations. Geometry of the quantum reduction of the
Dirac formalism to the physical sector of the theory is presented in
the coordinate gauges and extended to unitary momentum-dependent gauges of
a general type. The operators of physical observables are
constructed satisfying one-loop quantum gauge invariance and Hermiticity
with respect to a physical inner product. Abelianization procedure on
Lagrangian constraint surfaces of phase space is discussed in the framework
of the semiclassical expansion.
\end{abstract}

\section{Introduction}
\hspace{\parindent}
In this paper we discuss geometric properties of the quantum
dynamical systems subject to first class constraints
\cite{Dirac,DeWitt,Constr}. An importance of this problem arises in
view of the growing interest in quantization of the so-called zero modes
or moduli in various contexts of high energy physics ranging from theory
of gauge fields to quantum cosmology and nonperturbative D-brane aspects of
superstrings. In contrast with oscillatory degrees of freedom, quantization
of these modes is impossible in a conventional Fock space but, rather,
demands the coordinate representation of canonical commutation relations
\cite{Brink}. Quantum operators represented on configuration space of
the theory as differential operators have interesting geometric properties
with respect to diffeomorphisms of configuration space and group manifolds
and, thus, deserve careful analysis for the purpose of consistency,
uniqueness of quantization, and other subtle issues.

At the classical level systems subject to first class constraints are
described by the canonical action of generic form \cite{Constr}
        \begin{eqnarray}
        S=\int dt\,\{p_i\dot{q}^i-H_0(q,p)
        -\lambda^\mu T_\mu(q,p)\}                    \label{1.1}
        \end{eqnarray}
in the configuration space of coordinates and momenta $(q,p)=(q^i,p_i)$ and
Lagrange multiplyers $\lambda^\mu$. The variation of the latter leads to
the set of nondynamical equations -- constraints
        \begin{eqnarray}
        T_\mu(q,p)=0.                            \label{1.2}
        \end{eqnarray}
The first class constraint functions $T_\mu(q,p)$ and the Hamiltonian
$H_0(q,p)$ satisfy the Poisson-bracket algebra
        \begin{eqnarray}
        &&\{T_\mu,T_\nu\}=U^\lambda_{\mu\nu}T_\lambda,   \label{1.3}\\
        &&\{H_0,T_\nu\}=U^\lambda_{0\,\nu}T_\lambda,    \label{1.3a}
        \end{eqnarray}
with the structure functions $U^\lambda_{\mu\nu}=U^\lambda_{\mu\nu}(q,p)$
and $U^\lambda_{0\,\nu}=U^\lambda_{0\,\nu}(q,p)$ which can generally depend
on phase-space variables of the theory.

The first class constraints indicate that the theory possesses a local gauge
invariance generated in the sector of phase-space variables by
constraints themselves $T_\mu(q,p)$ and by certain
transformations of Lagrange multiplyers \cite{Constr}. The dimensionality
of the gauge group coincides with the dimensionality of the space of
constraints, the both being enumerated by the gauge index $\mu$. If we
denote the range of index $i$ by $n$, $i=1,...n$, and that of $\mu$ by $m$,
$\mu=1,...m$, then the number of the physical dynamically independent degrees
of freedom equals $n-m$: $2(n-m)$ physical phase-space variables originate
from the initial $2n$ variables $(q^i,p_i)$ by restricting to
$(2n-m)$-dimensional constraint surface (\ref{1.2}) and then factoring on this
surface out the action of $m$ gauge transformations generated by $T_\mu$.

Dirac quantization of the theory (\ref{1.1}) consists in promoting initial
phase-space variables and constraint functions to the
operator level $(q,p,T_\mu)\rightarrow (\hat{q},\hat{p},\hat{T}_\mu)$ and
selecting the physical states $|\,{\mbox{\boldmath$\Psi$}}\big>$ in the
representation space of $(\hat{q},\hat{p},\hat{T}_\mu)$ by the equation
        \begin{eqnarray}
        \hat{T}_\mu|\,{\mbox{\boldmath$\Psi$}}\big>=0.   \label{1.5}
        \end{eqnarray}
Operators $(\hat{q},\hat{p})$ satisfy canonical commutation relations
$[\hat{q}^k,\hat{p}_l]=i\hbar\delta^k_l$ and the quantum constraints
$\hat{T}_\mu$ as operator functions of $(\hat{q},\hat{p})$ should satisfy
the correspondence principle with classical $c$-number constraints
and be subject to the commutator algebra
        \begin{eqnarray}
        [\hat{T}_\mu,\hat{T}_\nu]=
        i\hbar\hat{U}^{\lambda}_{\mu\nu}\hat{T}_\lambda.  \label{1.6}
        \end{eqnarray}
with some operator structure functions $\hat{U}^{\lambda}_{\mu\nu}$ standing
to the left of operator constraints. This algebra generalizes (\ref{1.3}) to
the quantum level and serves as integrability conditions for equations
(\ref{1.5}).

Classically the theory (\ref{1.1}) and this reduction to its physical
sector has two types of invariances: the invariance with respect to
canonical transformations of the initial phase-space variables and the
geometric invariance with respect to the transformations of the basis of
constraints. The latter property means that one and the same constraint
surface (\ref{1.2}) is determined not just by one specific choice of the
set of constraint functions $T_\mu(q,p)$, but by the equivalence class of
those differing from one another by linear recombinations
        \begin{eqnarray}
        T_\mu^{\prime}=\Omega^{\,\nu}_{\,\mu}\, T_\nu,\,\,\,
        \Omega^\nu_\mu=\Omega^\nu_\mu(q,p),\,\,\,
        {\rm det}\,\Omega^\nu_\mu \neq 0       \label{1.7}
        \end{eqnarray}
with arbitrary invertible matrix function of $(q,p)$ acting in the vector
space of gauge indices. A natural question arises whether these invariances
can be preserved also in the Dirac quantization procedure?

The program of finding such quantum constraints for a generic
constrained system has been partly implemented in \cite{BKr} in the lowest
nontrivial order of semiclassical expansion in $\hbar$. The symbols of
operators $(\hat{T}_\mu,\,\hat{U}^{\lambda}_{\mu\nu})$ with linear in $\hbar$
quantum corrections have been found and a partial answer was given to the
question of the above type: the obtained operators turned out to be covariant
with respect to contact canonical transformations of the initial phase space,
provided the Dirac wavefunctions $\big<\,q\,|\,{\mbox{\boldmath$\Psi$}}\big>=
\mbox{\boldmath$\Psi$}(q)$ in the coordinate representation of
commutation relations for $(\hat{q},\hat{p})$ transform as 1/2-weight
densities on the configuration-space manifold of $q^i$.

It turned out, however, that the operator algorithms for
$(\hat{T}_\mu,\,\hat{U}^{\lambda}_{\mu\nu})$ involve not only their classical
counterparts featuring in (\ref{1.3}) but also the higher-order structure
functions of the canonical gauge algebra \cite{BV,BFV}. Generally the gauge
algebra involves the whole hierarchy of structure functions and relations
which begin with $T_\mu(q,p)$ and (\ref{1.3})
        \begin{eqnarray}
        G=\{T_\mu,U^{\alpha}_{\mu\nu},
        U^{\alpha\beta}_{\mu\nu\lambda},...\}   \label{1.8}
        \end{eqnarray}
and at any new stage iteratively
build up as consistency conditions for those of the previous stages. For
example, the cyclic Jacobi identity
$\{T_\mu,\{T_\sigma,T_\lambda\}\}+{\rm cycle}(\mu,\sigma,\lambda)=0$
applied to \ref{1.3} results in the equation
        \begin{eqnarray}
        \{T_\mu,U^\alpha_{\sigma\lambda}\}+
        U^\beta_{\sigma\lambda}U^\alpha_{\mu\beta}
        +{\rm cycle}(\mu,\sigma,\lambda)=
        U^{\alpha\beta}_{\mu\nu\lambda}T_\beta,      \label{1.9}
        \end{eqnarray}
multiplied by $T_\alpha$, necessarily generating a new structure function
$U^{\alpha\beta}_{\mu\nu\lambda}$ antisymmetric in upper (and lower) indices
\cite{BFV}. For constraints forming the closed Lie algebra all higher-order
structure functions are vanishing, but this property depends on the
choice of basis of constraints: the rotation of the constraint basis
(\ref{1.7}) can convert the Lie algebra (even Abelian one with
$U^\alpha_{\sigma\lambda}=0$) into an open algebra with the infinite set
of structure functions. Thus the invariance of the the theory with respect to
transformations of the form (\ref{1.7}) with arbitrary $\Omega^\mu_\nu$
necessitates considering higher-order structure functions (\ref{1.8})
and their operator realization $G\rightarrow \hat{G}=\{\hat{T}_\mu,
\hat{U}^{\alpha}_{\mu\nu},\hat{U}^{\alpha\beta}_{\mu\nu\lambda},...\}$.

In this paper we show that the operators constructed in \cite{BKr} really
possess the expected properties of invariance with respect to the
transformation of the constraint basis (\ref{1.7}) -- the result briefly
reported earlier in \cite{geom}. Since we restrict ourselves with the
one-loop (linear in $\hbar$) approximation, we shall focuse at the
covariance of only the Dirac constraints $\hat{T}_\mu$ and Dirac equations
on physical states (\ref{1.5}): the higher-order structure functions
will not be important for us because they are responsible for multi-loop
orders of the semiclassical expansion.

The paper is organized as follows. In Sect.2 we extend the operator
realization of quantum constraints of \cite{BKr} to physical observables.
Sect. 3 contains the proof of the covariance of equations (\ref{1.5}), which
induces the weight properties of Dirac wavefunctions in
the space of gauge indices. These properties guarantee the unitary
equivalence of quantum theories starting with different choices of constraint
bases, equipped with a correct physical inner product. Sect. 4 gives the
the fundamental two-point solution of quantum Dirac constraints explicitly
featuring the weight properties of the above type. The reduction of this
solution to physical sector is performed in Sect. 5, where this reduction
(previously known only for a narrow class of {\em coordinate} gauges
\cite{GenSem,BKr,geom}) is extended to unitary gauges of general type.
In Sect.6 we consider a number of other issues omitted in paper \cite{BKr}:
operator
gauge independence of matrix elements of physical observables and their
Hermitian conjugation properties in the physical inner product. The concluding
section contains a brief discussion of possible applications of these results,
while two appendices give a number of technical details regarding the
Hermiticity of observables and an abelianization procedure for
semiclassical constrained systems on Lagrangian manifolds of phase space.

\section{Operator realization of quantum constraints and physical observables}
\hspace{\parindent}
The operator realization of quantum Dirac constraints and lowest order
structure functions was found in \cite{BKr} in the form of the normal
$qp$-ordering of their $qp$-symbols expanded up to the linear order in
$\hbar$. This
representation implies that for any operator $\hat{G}=\{\hat{T}_\mu,
\hat{U}^{\alpha}_{\mu\nu},\hat{U}^{\alpha\beta}_{\mu\nu\lambda},...\}$ one
can put into
correspondence its normal symbol -- a $c$-number function on phase space
$\tilde{G}(q,p)$ -- such that the operator $\hat{G}$ can be obtained from
$\tilde{G}(q,p)$ by replacing its arguments with noncommuting operators
with all the momenta standing to the right of coordinates. For a symbol
expandable in momentum series
        \begin{eqnarray}
        \tilde{G}(q,p)=\sum_{n=0}^{\infty}\tilde{G}^{i_1 ...i_n}(q)
        p_{i_1}...p_{i_n}                         \label{2.1}
        \end{eqnarray}
this means that
        \begin{eqnarray}
        \hat{G}={\cal N}_{qp}\tilde{G}(q,p)\equiv
        \sum_{n=0}^{\infty}\tilde{G}^{i_1 ...i_n}(\hat{q})
        \hat{p}_{i_1}...\hat{p}_{i_n}.            \label{2.2}
        \end{eqnarray}

The one-loop (linear in $\hbar$) algorithms of \cite{BKr} for two lowest-order
operators $\hat{G}$ have the form
        \begin{eqnarray}
        &&\hat{T}_{\mu}={\cal N}_{qp}\left\{T_{\mu}-
        \frac{i\hbar}2\frac{\partial^2\, T_{\mu}}{\partial q^i
        \partial p_i}+\frac{i\hbar}2 U^{\nu}_{\mu\nu}+
        O(\hbar^2)\right\},                     \label{2.3}\\
        &&\hat{U}^{\lambda}_{\mu\nu}=
        {\cal N}_{qp}\left\{U^{\lambda}_{\mu\nu}-
        \frac{i\hbar}2\frac{\partial^2\,
        U^{\lambda}_{\mu\nu}}{\partial q^i
        \partial p_i}-\frac{i\hbar}2 U^{\lambda\sigma}_{\mu\nu\sigma}+
        O(\hbar^2)\right\}                     \label{2.4}
        \end{eqnarray}
involving, as it was mentioned in Introduction, the higher-order classical
structure functions $U^{\lambda\sigma}_{\mu\nu\sigma}$. As shown in
\cite{BKr},
these operators have two important properties. First, they are covariant
under contact canonical transformations of $(q,p)$
        \begin{eqnarray}
        &&q^i=q^i(q^{\prime}),\,\,\,p_i=p_{k^\prime}
        \frac{\partial q^{k^\prime}}{\partial q^i},
        \,\,\, G(q,p)=G'(q',p'),                      \label{2.5}
        \end{eqnarray}
under which the constraints and structure functions (\ref{1.8}) (all
quantities bearing only gauge indices) behave like scalars.
In the coordinate representation of canonical commutation relations the
covariance of operators (\ref{2.3})-(\ref{2.4}) can be written down as
        \begin{eqnarray}
        \left|\frac{\partial q^\prime}{\partial q}\right|^{-1/2}
        \hat{G}\,\left|\frac{\partial q^\prime}
        {\partial q}\right|^{1/2}
        =\hat{G}^\prime,                           \label{2.6}
        \end{eqnarray}
where the operators $\hat{G}^\prime$ are constructed by the above algorithms
from their primed classical counterparts. This transformation law obviously
implies that the Dirac wavefunction satisfying quantum constraints (\ref{1.5})
should be regarded a scalar density of 1/2-weight
        \begin{eqnarray}
        \mbox{\boldmath$\Psi$}(q)=
        \left|\frac{\partial q^\prime}{\partial q}\right|^{1/2}
        \mbox{\boldmath$\Psi$}'(q')                 \label{2.7}
        \end{eqnarray}
in complete correspondence with the diffeomorphism invariance of the
{\em auxiliary} inner product of {\em unphysical} states
        \begin{eqnarray}
        \big<{\mbox{\boldmath$\Psi$}}_1
        |{\mbox{\boldmath$\Psi$}}_2\big>=\int dq\,
        {\mbox{\boldmath$\Psi$}}_1^*(q)
        {\mbox{\boldmath$\Psi$}}_2(q).             \label{2.8}
        \end{eqnarray}
This inner product diverges for physical states satisfying quantum Dirac
constraints and, therefore, plays only an auxiliary role. It appears
as a truncated inner product in the bosonic sector of the extended
configuration space of the BFV (BRST) quantization
\cite{BFV,BKr,Marn1,Marn2,BatMarn} and also will be used below (in Sect.6)
for the construction of the {\em physical} inner product by means of a
special operator measure.

The second important property of these operators with respect to the inner
product (\ref{2.8}) is their anti-Hermitian part. It is given by the trace
of the structure functions and for Dirac constraints has the form
        \begin{eqnarray}
        \hat{T}_\mu-\hat{T}_\mu^\dagger=
        i\hbar(\hat{U}^\lambda_{\mu\lambda})^\dagger
        +O(\hbar^2).                               \label{2.9}
        \end{eqnarray}

The algorithms (\ref{2.3})-(\ref{2.4}) were derived in \cite{BKr} solely
as a solution of the commutator algebra (\ref{1.6}). This helps to extend
these algorithms for obtaining another class of operators -- the operators of
physical observables. Classically the physical observables ${\cal O}_I$
(enumerated by some index $I$) are
defined as a functions on phase space, invariant under the action of
canonical gauge algebra. This invariance generally holds in a weak sense,
that is only on the constraint surface
        \begin{eqnarray}
        \{{\cal O}_I,T_\mu\}=U^\lambda_{I\,\mu}T_\lambda,  \label{2.10}
        \end{eqnarray}
the gauge transformation of ${\cal O}_I$ being a linear combination
of constraints with some coefficients $U^\lambda_{I \mu}=
U^\lambda_{I \mu}(q,p)$. ($H_0(q,p)$ and $U^\lambda_{0\,\mu}$ with
eq.(\ref{1.3a}) present the example of such an observable and its weak
invariance.) Note that again due to the rotation of the
constraint basis (\ref{1.7}) we have to consider nonvanishing
coefficients $U^\lambda_{I\,\mu}$ which can always be generated even for
strongly invariant observables by a transition to another basis of
constraints.

In addition to their weak invariance (\ref{2.10}) we shall assume that
the classical observables {\em commute} with one another or form
{\em a closed Lie} algebra in a weak sense
        \begin{eqnarray}
        \{{\cal O}_I,{\cal O}_J\}=U^L_{IJ}{\cal O}_L
        +U^\lambda_{IJ}T_\lambda,\,\,\,U^L_{IJ}={\rm const}. \label{2.14}
        \end{eqnarray}
In this case, from the viewpoint of commutator algebra the physical
observables do not differ from constraints. The only difference is that unlike
constraints they are not constrained to vanish. Therefore, to promote
the classical observables to the quantum level,
$({\cal O}_I,U^\lambda_{I\,\mu})
\rightarrow (\hat{\cal O}_I,\hat{U}^\lambda_{I\,\mu})$, and enforce
the quantum gauge invariance of their operators
        \begin{eqnarray}
        [\hat{\cal O}_I,\hat{T}_\mu]
        =i\hbar\hat{U}^\lambda_{I\,\mu}\hat{T}_\lambda,   \label{2.11}
        \end{eqnarray}
one can use the algorithm analogous to (\ref{2.3}) solving this commutator
algebra
        \begin{eqnarray}
        &&\hat{\cal O}_I={\cal N}_{qp}\left\{{\cal O}_I-
        \frac{i\hbar}2\frac{\partial^2\, {\cal O}_I}{\partial q^i
        \partial p_i}+\frac{i\hbar}2 U^{\lambda}_{I\,\lambda}+
        \frac{i\hbar}2 U^J_{IJ}+
        O(\hbar^2)\right\},                             \label{2.12}\\
        &&\hat{U}^{\lambda}_{I\,\mu}=
        {\cal N}_{qp}\left\{U^{\lambda}_{I\,\mu}-
        \frac{i\hbar}2\frac{\partial^2\,
        U^{\lambda}_{I\,\mu}}{\partial q^i
        \partial p_i}-\frac{i\hbar}2 U^{\lambda\sigma}_{I\,\mu\sigma}+
        O(\hbar^2)\right\}                             \label{2.13}
        \end{eqnarray}
with higher-order structure functions $U^{\lambda\sigma}_{I\,\mu\sigma}$
of the classical algebra (\ref{2.10}) and (\ref{2.14}) (derivable by the
method mentioned in Introduction).

The quantum observables (\ref{2.12}) solve the closed commutator
algebra (\ref{2.11}). The proof of this statement goes by collecting the
observables together with constraints into one set and repeating the
derivation of \cite{BKr}. The only thing to check is if the resulting
commutator
algebra does not contain nonvanishing components $\hat{U}^K_{I\,\mu}$
of the operator structure functions (violating the weak quantum gauge
invariance of observables (\ref{2.11})). This component can get a
nonvanishing contribution only due to a higher-order structure functions
$U^{K \lambda}_{I \mu\lambda}+U^{KL}_{I \mu L}$ of the classical algebra
(\ref{2.10}) and (\ref{2.14}). But it is easy to show that for a {\em closed
Lie} algebra (\ref{2.14}) the nonvanishing components of the
second-order structure functions cannot have {\em nongauge} upper indices
($U^{J\,\lambda}_{\,...}=0,\,\,U^{JL}_{\,...}=0$) and, therefore the
quantum observables remain weakly invariant. For the same reason
eq.(\ref{2.13}) does not involve the contraction
$U^{\lambda\,J}_{I\,\mu\,J}$.

The quantum observables like constraints have anti-Hermitian part with
respect to the auxiliary inner product (\ref{2.8}). It is given by
two contractions of structure functions $U^{\lambda}_{I\,\lambda}$
and $U^J_{IJ}$. For all reasonable compact groups generating algebras of
observables the latter is vanishing $U^J_{IJ}=0$, but
$U^{\lambda}_{I\,\lambda}$ is generally nontrivial, depends on the choice of
constraint basis and violates Hermeticity of observables in the auxiliary
inner product. It is however inessential, because only the physical inner
product must generate real expectation values of observables, and this
will be shown to be true below.

\section{Quantum transformation of the constraint basis}
\hspace{\parindent}
Under the linear transformation of the classical constraint basis (\ref{1.7})
the structure functions transform as
        \begin{eqnarray}
        U'^\sigma_{\mu\nu}=\Omega^\alpha_\mu \Omega^\beta_\nu
        U^\lambda_{\alpha\beta}\Omega^{-1\,\sigma}_{\,\,\,\lambda}+
        2\{\Omega^\alpha_{[\mu},T_\beta\} \Omega^\beta_{\nu]}
        \Omega^{-1\,\sigma}_{\,\,\,\alpha}+\{\Omega^\alpha_\mu,
        \Omega^\beta_\nu\} T_\alpha
        \Omega^{-1\,\sigma}_{\,\,\,\beta},     \label{3.1}
        \end{eqnarray}
so that
        \begin{eqnarray}
        &&U'^\lambda_{\mu\lambda}=\Omega^\alpha_\mu
        U^\lambda_{\alpha\lambda}+\{\Omega^\alpha_\mu,T_\alpha\}
        -\{\ln \Omega,\Omega^\alpha_\mu T_\alpha\},       \label{3.2}\\
        &&\Omega\equiv\det \Omega^\alpha_\beta.           \label{3.3}
        \end{eqnarray}

The quantum constraints $\hat{T}'_\mu$ based on the transformed basis of
classical constraints (\ref{1.7}) and structure functions (\ref{3.1}) take
on the use of the algorithm (\ref{2.3}) (with primed quantities) the form
        \begin{eqnarray}
        \hat{T}'_\mu={\cal N}_{qp}\left[\tilde{\Omega}^\nu_\mu
        \tilde{T}_\nu-i\hbar\frac{\partial\tilde{\Omega}^\nu_\mu}
        {\partial p_k}\frac{\partial\tilde{T}_\nu}{\partial q^k}-
        \frac{i\hbar}2\{\ln\tilde{\Omega},\tilde{\Omega}^\nu_\mu
        \tilde{T}_\nu\}+O(\hbar^2)\right],                   \label{3.4}
        \end{eqnarray}
where $\tilde{T}_\nu$ is a normal $qp$-symbol of constraints in the original
basis and
        \begin{eqnarray}
        \tilde{\Omega}^\nu_\mu\equiv
        \Omega^\nu_\mu-\frac{i\hbar}2\frac{\partial^2\,
        \Omega^\nu_\mu}{\partial q^k
        \partial p_k}+O(\hbar^2),\,\,\,\,
        \tilde{\Omega}\equiv\det\tilde{\Omega}^\nu_\mu.   \label{3.5}
        \end{eqnarray}

From the two simple identities for operators and their normal $qp$-symbols
        \begin{eqnarray}
        &&\hat{F}_1\hat{F}_2={\cal N}_{qp}\left[\tilde{F}_1\tilde{F}_2
        -i\hbar\frac{\partial\tilde{F}_1}{\partial p_k}
        \frac{\partial\tilde{F}_2}{\partial q^k}
        +O(\hbar^2)\right],                         \nonumber\\
        &&\hat{\Omega}^{-1/2}\hat{F}\hat{\Omega}^{1/2}=
        {\cal N}_{qp}\left[\tilde{F}-
        \frac{i\hbar}2\{\ln \tilde{\Omega},\tilde{F}\}
        +O(\hbar^2)\right]                         \nonumber
        \end{eqnarray}
it is then easy to find the final form of the transformation law for
quantum constraints under the transformation of their classical basis
(\ref{1.7})
        \begin{eqnarray}
        &&\hat{T}'_\mu=\hat{\Omega}^{-1/2}
        \hat{\Omega}^\nu_\mu\hat{T}_\nu
        \hat{\Omega}^{1/2},                     \label{3.6}\\
        &&\hat{\Omega}^\nu_\mu={\cal N}_{qp}\left\{\Omega^\nu_\mu
        -\frac{i\hbar}2\frac{\partial^2\,
        \Omega^\nu_\mu}{\partial q^k
        \partial p_k}+O(\hbar^2)\right\}, \,\,\,\,
        \hat{\Omega}=\det\hat{\Omega}^\mu_\nu.     \label{3.7}
        \end{eqnarray}

Similarly to (\ref{1.7}) this transformation involves a linear recombination
of constraint operators with operator-valued matrix $\hat{\Omega}^\nu_\mu$
(standing to the left of constraints). This matrix is obtained from
its classical counterpart $\Omega^\mu_\nu(q,p)$ by the algorithm (\ref{3.7})
similar to (\ref{2.3}): its symbol involves analogous quantum corrections
except the anti-Hermitian part. In addition to linear combinations the
transformation (\ref{3.6}) includes the
canonical transformation generated by the square root of its determinant
$\hat{\Omega}^{1/2}$. This canonical transformation implies that the physical
states satisfying Dirac constraints (\ref{1.5}) transform contragrediently
to (\ref{3.6})
        \begin{eqnarray}
        |\,{\mbox{\boldmath$\Psi$}}\big>'=\hat{\Omega}^{-1/2}
        |\,{\mbox{\boldmath$\Psi$}}\big>                 \label{3.8}
        \end{eqnarray}
and turn out to be scalar densities of weight $-1/2$ in the space of gauge
indices. This property has been observed for systems subject to constraints
linear in momenta in \cite{FHenneauxP} and, as we see, turns out to be true
for a generic case at least in the one-loop order of the semiclassical
expansion.

It is easy to repeat now similar calculations for the operators of physical
observables (\ref{2.12}). In view of the transformation law for the trace
of the observable structure function $U'^\lambda_{I\,\lambda}=
U^\lambda_{I\,\lambda}-\{\ln\det\Omega,{\cal O}_I\}$ these calculations
immeadiately show that the operators of observables also transform
canonically
        \begin{eqnarray}
        \hat{\cal O}'_I=\hat{\Omega}^{-1/2}\hat{\cal O}_I
        \hat{\Omega}^{1/2}.                             \label{3.8a}
        \end{eqnarray}

Obviously, the theories differing by the choice of constraint basis should
be unitarily equivalent. This means that the physical inner product of
states $|\,{\mbox{\boldmath$\Psi$}}\big>$ should contain a measure
depending on this choice and transforming contragrediently to (\ref{3.8}).
In the next section we show that semiclassical states and their inner product
really satisfy the properties compatible with the transformation law of the
above type.

\section{Semiclassical physical states}
\hspace{\parindent}
Semiclassical expansion of the operator symbols of the above type makes
sense when the corresponding quantum states also have a semiclassical
form. In the coordinate representation semiclassical wavefunctions
        \begin{eqnarray}
        {\mbox{\boldmath$\Psi$}}(q)={\mbox{\boldmath$P$}}(q)
        \exp\left[\frac i{\hbar}
        {\mbox{\boldmath$S$}}(q)\right]           \label{4.1}
        \end{eqnarray}
are characterized by the Hamilton-Jacobi function ${\mbox{\boldmath$S$}}(q)$
and preexponential factor ${\mbox{\boldmath$P$}}(q)$ expandable in
$\hbar$-series beginnig with the one-loop order\footnote
{This corresponds to the fact that the tree-level part is entirely contained
in the exponential and is $O(\hbar^{-1})$. When the Hamiltonian $H_0$
in eq.(\ref{1.1}) is nonvanishing the wavefunction
(\ref{4.1}), its Hamilton-Jacobi function, two-point kernel, etc. are
time-dependent. In what follows we shall, however, omit the time label,
because we will be mainly interested in constraint properties rather than
the dynamical ones. Another way to view this is to parametrize time and the
conjugated Hamiltonian $H_0(q,p)$ among the phase-space variables and regard
the Hamilton-Jacobi and Schrodinger equations as one extra classical and
quantum constraint correspondingly \cite{BKr,BarvU}.
}
$O(\hbar^0)$. The action of the operator $\hat{F}$ on such functions
reads as
        \begin{eqnarray}
        \hat{F}{\mbox{\boldmath$\Psi$}}(q)=
        \left[\tilde{F}\left(q,\frac{\partial{\mbox{\boldmath$S$}}}
        {\partial q}\right)+O(\hbar)\right]
        {\mbox{\boldmath$P$}}
        \exp\left[\frac i{\hbar}
        {\mbox{\boldmath$S$}}(q)\right]              \label{4.2}
        \end{eqnarray}
where, as in (\ref{2.2}), $\tilde{F}$ is a normal $qp$-symbol of $\hat{F}$.

The general semiclassical solution of quantum constraints (\ref{1.5}) with
operators (\ref{2.3}) was found in \cite{GenSem,BKr,BarvU} in the form of the two-point
kernel ${\mbox{\boldmath$K$}}(q,q')$ "propagating" the initial data
from the Cauchy surface throughout the whole superspace of $q$
        \begin{eqnarray}
        {\mbox{\boldmath$K$}}(q,q')={\mbox{\boldmath$P$}}(q,q')
        \exp\left[\frac i{\hbar}
        {\mbox{\boldmath$S$}}(q,q')\right].               \label{4.3}
        \end{eqnarray}
In both expressions (\ref{4.1}) and (\ref{4.3}) the phase in the exponential
satisfies the Hamilton-Jacobi equation
        \begin{eqnarray}
        T_\mu\left(q,\frac{\partial{\mbox{\boldmath$S$}}}
        {\partial q}\right)=0,                        \label{4.4}
        \end{eqnarray}
while the one-loop preexponential factor is subject to continuity type
equation originating from the full quantum constraint in the approximation
linear in $\hbar$
        \begin{eqnarray}
        &&\frac\partial{\partial q^i}
        (\nabla^i_\mu {\mbox{\boldmath$P$}}^2)=
        U^{\lambda}_{\mu\lambda}
        {\mbox{\boldmath$P$}}^2,                      \label{4.5}\\
        &&\nabla^i_\mu\equiv\left.
        \frac{\partial T_{\mu}}{\partial p_i}
        \right|_{\textstyle p=\partial
        {\mbox{\boldmath$S$}}/\partial q}.            \label{4.6}
        \end{eqnarray}

For a two-point kernel the Hamilton-Jacobi function coincides with the
principal Hamilton function ${\mbox{\boldmath$S$}}(q,q')$ (action on the
extremal joining points $q$ and
$q'$) and the solution of the continuity equation can be found as a
generalization of the Pauli-Van Vleck-Morette ansatz for the one-loop
preexponential factor \cite{Cecile,BarvU} of the Schrodinger propagator. This
generalization is nothing but a Faddeev-Popov gauge-fixing \cite{FP}
procedure for a matrix of mixed second-order derivatives of the principal
Hamilton function
        \begin{eqnarray}
        {\mbox{\boldmath$S$}}_{ik'}=
        \frac{\partial^2{\mbox{\boldmath$S$}}(q,q')}
        {\partial q^i\, \partial q^{k'}}             \label{4.7}
        \end{eqnarray}
which is degenerate in virtue of the Hamilton-Jacobi equations (\ref{4.4})
giving rise to the left zero-value eigenvectors (\ref{4.6}) and analogous
right zero-vectors \cite{GenSem,BKr}
        \begin{eqnarray}
        \nabla^i_\mu {\mbox{\boldmath$S$}}_{ik'}=0,\,\,\,
        {\mbox{\boldmath$S$}}_{ik'}\nabla^{k'}_\nu=0,\,\,\,
        \nabla^{k'}_\nu\equiv\left.
        \frac{\partial T_{\nu}(q',p')}{\partial p'_{k}}
        \right|_{\textstyle p'=-\partial
        {\mbox{\boldmath$S$}}/\partial q'}.   \label{4.8}
        \end{eqnarray}

The preexponential factor reads
        \begin{eqnarray}
        {\mbox{\boldmath$P$}}=
        \left[\frac{{\rm det}{\mbox{\boldmath$F$}}_{ik'}}
        {J(q)J(q')\,{\rm det}\,c_{\mu\nu}}\right]^{1/2},           \label{4.9}
        \end{eqnarray}
where ${\mbox{\boldmath$F$}}_{ik'}$ is a nondegenerate matrix of the initial
action Hessian (\ref{4.7}) supplied with a gauge-breaking term
        \begin{eqnarray}
        {\mbox{\boldmath$F$}}_{ik'}=
        {\mbox{\boldmath$S$}}_{ik'}+
        \chi^\mu_i c_{\mu\nu} \chi^\nu_{k'},             \label{4.10}
        \end{eqnarray}
and $J(q)$ and $J(q')$ are the Feynman-DeWitt-Faddeev-Popov "ghost"
determinants \cite{DW,FP} compensating for the inclusion of this term.
The gauge-breaking term and
ghost determinants are constructed with the aid of two sets of arbitrary
covectors $(\chi^\mu_i,\chi^\nu_{k'})$ ("gauge" conditions) satisfying the
only requirement of the nondegeneracy of their ghost operators
\cite{GenSem,BarvU,BKr}
        \begin{eqnarray}
        &&J^\mu_\nu(q)=\chi^\mu_i\nabla^i_\nu,\,\,
        J(q)\equiv{\rm det}J^\mu_\nu(q)\neq 0,\nonumber\\
        &&J^{\mu}_{\nu}(q')=\chi^\mu_{i'}\nabla^{i'}_\nu,\,\,
        J(q')\equiv{\rm det}J^{\mu}_{\nu}(q')\neq 0.       \label{4.11}
        \end{eqnarray}
The invertible gauge-fixing matrix $c_{\mu\nu}$ and its determinant
(contribution of Nielsen-Kallosh ghosts) are the last ingredients of
the generalized Pauli-Van Vleck-Morette ansatz (\ref{4.9}).

Notice now that under the transformation of the basis (\ref{1.7}) the vectors
(\ref{4.6}) defined on the Lagrangian manifold of phase-space
$p=\partial{\mbox{\boldmath$S$}}/\partial q$ transform covariantly with
respect to their gauge indices $(\nabla^i_\mu)'=\Omega^\nu_\mu \nabla^i_\nu$.
Therefore the ghost determinants transform as densities
        \begin{eqnarray}
        J'=(\det \Omega^\mu_\nu)J,                         \label{4.12}
        \end{eqnarray}
whence it follows that the two-point kernel with respect to each of its
arguments transforms in accordance with the law (\ref{3.8}) in which
the action of the operator $\hat{\Omega}^{-1/2}$ semiclassically boils
down to the multiplication with $\left[\det \Omega^\mu_\nu
(q,\partial{\mbox{\boldmath$S$}}/\partial q)\right]^{-1/2}$.

\section{Reduction to the physical sector: unitary gauge conditions of
a general type}
\hspace{\parindent}
The interpretation of the semiclassical state (\ref{4.3}), (\ref{4.9}) is
rather transparent in the physical sector of the theory
\cite{GenSem,BKr,BarvU}. The sector explicitly arises after the reduction
to physical variables by disentangling them from the original phase space
of $(q^i,p_i)$ in a unitary gauge. Such a reduction in
\cite{GenSem,BKr,BarvU} was given for a special type of gauge conditions
imposed only on phase space coordinates $q^i$. Here we generalize this
reduction procedure to unitary gauges of a general type.

The theory with the action (\ref{1.1}) is invariant under the set of
canonical transformations of phase space variables generated by constraints
and special (noncanonical) transformations of Lagrange multiplyers
\cite{Constr,BarvU}
        \begin{eqnarray}
        &&\delta^{\cal F}q^i=\{q^i,T_\mu\}{\cal F}^\mu,\,\,\,
        \delta^{\cal F}q^i=\{q^i,T_\mu\}{\cal F}^\mu, \label{5.01}\\
        &&\delta^{\cal F}\lambda^\mu=\dot{\cal F}^\mu
        -U^\mu_{\nu\lambda}\lambda^\nu{\cal F}^\lambda
        -U^\mu_{0\,\lambda}{\cal F}^\lambda            \label{5.02}
        \end{eqnarray}
with arbitrary infinitesimal gauge parameter ${\cal F}^\mu={\cal F}^\mu(t)$.
This invariance implies that the equivalence class of variables belonging to
the orbit of these transformations corresponds to one and the same physical
state. The description of this state in terms of physical variables rather
than the equivalence classes of original degrees of freedom consists in
singling out the unique representative of each such class and treating the
independent labels of this representative as physical variables. This can be
attained by imposing on original phase space variables the gauge conditions
        \begin{eqnarray}
        \chi^\mu(q,p)=0                      \label{5.03}
        \end{eqnarray}
which determine in the $2n$-dimensional phase space the $(2n-m)$-dimensional
surface (remember that $n$ is the range of index $i$, while $m$ is that
of $\mu$) having a unique intersection with the orbit of gauge
transformations (\ref{5.01}). At least locally, the latter condition means
the invertibility of the Faddeev-Popov matrix with the nonvanishing
determinant
        \begin{eqnarray}
        J(q,p)=\det J^\mu_\nu(q,p),\,\,\,J^\mu_\nu(q,p)=
        \{\chi^\mu,T_\nu\}.                              \label{5.2}
        \end{eqnarray}
Gauge conditions of the form (\ref{5.03}) are called unitary. They impose
restrictions only on phase space variables and, thus, locally in time single
out the physical sector and allow one to formulate the gauge theory in a
manifestly unitary form. In this respect they differ from nonunitary gauge
conditions imposed on Lagrange multiplyers $\lambda^\mu$ (together with
$(q,p)$) and/or their time derivatives \cite{Constr,BarvU}.

\subsection{Coordinate gauge conditions}
\hspace{\parindent}
Unitary gauge conditions take the simplest form when they are imposed only on
phase space coordinates, $\chi^\mu(q)=0$. Such {\em coordinate} gauge
conditions determine the embedding of the $(n-m)$-dimensional
space $\Sigma$ of physical {\em coordinates} directly into the
space of original coordinates $q^i$ -- superspace. This fact
strongly simplifies the reduction of the semiclassical kernel (\ref{4.3}),
(\ref{4.9}) to the physical sector, because this reduction in
the main boils down to the embedding of the arguments of
${\mbox{\boldmath$K$}}(q,q')$ into the physical subspace $\Sigma$.
The geometry of this embedding, considered in much detail in
\cite{BarvU}, can be better described in special coordinates on superspace
$\bar{q}^i=(\xi^A,\theta^\mu)$, in which $\xi^A,\, A=1,...n-m,$ serve as
intrinsic coordinates on $\Sigma$ (physical configuration coordinates),
and $\theta^\mu$ is determined by gauge conditions:
        \begin{eqnarray}
        q^i\rightarrow \bar{q}^i=(\xi^A,\theta^\mu),\,\,
        q^i=e^i(\xi^A,\theta^\mu),\,\,
        \theta^\mu=\chi^\mu(q).         \label{5.3}
        \end{eqnarray}
The equation of the surface $\Sigma$ in the new coordinates is $\theta^\mu=0$,
so that its embedding equations coincide with the above reparametrization
equations at $\theta^\mu=0,\,\, e^i(\xi)=e^i(\xi,0)$
        \begin{eqnarray}
        \Sigma: q^i=e^i(\xi),\,\,\,
        \chi^\mu(e^i(\xi))\equiv 0.   \label{5.4}
        \end{eqnarray}
The relation between the integration measures on superspace $dq=d^n q$ and
on $\Sigma$, $d\xi=d^{n-m}\xi$
        \begin{eqnarray}
        d\xi=dq\,\delta(\chi)\,M,\,\,\,
        M=({\rm det}[e^i_A,e^i_\mu])^{-1},      \label{5.5}
        \end{eqnarray}
involves the Jacobian of this reparametrization, built of the basis of vectors
tangential and normal to $\Sigma$:
        \begin{eqnarray}
        e^i_A=\partial e^i/\partial\xi^A,\,\,\,\,
        e^i_\mu=\partial e^i/\partial\theta^\mu.    \label{5.5a}
        \end{eqnarray}
Note that $m$ covectors normal to the surface can be chosen as gradients of
gauge conditions
        \begin{eqnarray}
        \chi^\mu_i=\frac{\partial \chi^\mu}{\partial q^i},\,\,\,
        \chi^\mu_i e^i_\nu=\delta^\mu_\nu,              \label{5.6}
        \end{eqnarray}
that can be identified with auxiliary covectors
participating in the algorithm for the preexponential factor (\ref{4.9}).
With this identification the Faddeev-Popov operator
$J^\mu_\nu(q,\partial{\mbox{\boldmath$S$}}/\partial q)$ coincides with the
operator $J^\mu_\nu(q)$ of this algorithm (which explains the use of the same
notation).

On the same footing with $(e^i_A,e^i_\mu)$ as a full local
basis one can also choose the set $(e^i_A,\nabla^i_\mu)$ with vectors
$\nabla^i_\mu$ transversal to $\Sigma$ given by eq.(\ref{4.6}). The
normal vectors of the first basis when expanded in the new basis
        \begin{eqnarray}
        e^i_\mu=J^{-1\,\nu}_{\,\,\,\,\,\mu}\nabla^i_\nu
        +\Omega^A_\mu e^i_A                            \label{5.7}
        \end{eqnarray}
have one expansion coefficient always determined by the inverse of the
Faddeev-Popov matrix $J^{-1\,\nu}_{\,\,\,\,\,\mu}$ and, thus, independent of
the particular parametrization of $\Sigma$ by internal coordinates. The second
coefficient is less universal and depends on a particular choice of this
parametrization. Missing information about $\Omega^A_\mu$ does not
prevent, however, from finding the relation between the determinants of
matrices of the old and new bases
        \begin{eqnarray}
        {\rm det}\,[e^i_A,\,\nabla^i_\mu]=\frac JM.    \label{5.8}
        \end{eqnarray}

The reduction to physical sector in coordinate gauges follows after
identifying $\xi^A$ with the physical coordinates. The corresponding
conjugated momenta $\pi_A$ can be found from the transformation of the
symplectic form restricted to the physical subspace (\ref{5.4})
        \begin{eqnarray}
        &&\int dt\,p_i\dot{q}^i=\int dt\,
        \left(p_i e^i_A\dot{\xi}^A
        +p_i\frac{\partial e^i(\xi,t)}{\partial t}\right),    \label{5.8a} \\
        &&\pi_A=p_i e^i_A,
        \end{eqnarray}
as projections of the original momentum to the tangential components of the
basis (\ref{5.5a}). The normal projections of $p_i$ should be found from
constraints (\ref{1.2}), the local uniqueness of their solution being granted
by the nondegeneracy of the Faddeev-Popov determinant. Together with
(\ref{5.4}) this solution yields all the original phase space variables
$(q^i,p_i)$ as known functions of the physical degrees of freedom
$(\xi^A,\pi_A)$. The original action (\ref{1.1}) reduced to physical sector
(that is to the subspace of constraints and gauge conditions) acquires the
usual canonical form with the physical Hamiltonian contributed by the second
term of (\ref{5.8a}) and $H_0(q,p)$. Note that generically, especially for
the so-called parametrized systems with $H_0(q,p)=0$ in (\ref{1.1}),
canonical gauge conditions should explicitly depend on time,
$\chi^\mu(q)=\chi^\mu(q,t)$, in order to generate the dynamical evolution in
reduced phase space theory \cite{BarvU,BKr}. Therefore, the reduced
symplectic form generates a nontrivial contribution to the physical
Hamiltonian, proportional to the time derivative of the embedding
functions (\ref{5.4}) explicitly depending on $t$, $q=e(\xi,t)$. The total
physical Hamiltonian then takes the form
        \begin{eqnarray}
        H_{\rm red}(\xi,\pi)=\left[\,H_0(q,p)
        -p_i\frac{\partial e^i(\xi,t)}{\partial t}\,
        \right]_{q=q(\xi,\pi),\,p=p(\xi,\pi)}.
        \end{eqnarray}

Canonical quantization of such a classical system runs as usual along
the lines of a particular representation and operator realization in the
Hilbert space of the theory.
In the one-loop (linear in $\hbar$) approximation with the Weyl ordering of
the above Hamiltonian this quantization is basically exhausted by the unitary
evolution kernel $K(t,\xi|t',\xi')$ of the Schrodinger equation. In the
coordinate representation it is given by the well-known Pauli-Van
Vleck-Morette ansatz \cite{Cecile}
        \begin{eqnarray}
        &&K(t,\xi|t',\xi')\equiv
        \left[{\rm det}\,\frac i{2\pi\hbar}
        \frac{\partial^2 S(t,\xi|t',\xi')}
        {\partial\xi^A\,\partial\xi^{B'}}\right]^{1/2}
        \,e^{\,{}^{\textstyle{\frac i{\hbar}
        S(t,\xi|t',\xi')}}},                          \label{5.10}
        \end{eqnarray}
which involves the principal Hamilton function of physical variables
$S(t,\xi|t',\xi')$ -- a classical action evaluated at the classical
extremal passing the points $\xi'$ and $\xi$ respectively at initial $t'$
and final $t$ moments of time. The preexponential factor here is built of
Van-Vleck determinant and guarantees in the approximation linear in $\hbar$
the unitarity of the Schrodinger evolution of the physical states
$\Psi(t,\xi)$
        \begin{eqnarray}
        \Psi(t,\xi)=\int d\xi'\,K(t,\xi|t',\xi')\,
        \Psi(t',\xi')                                \label{5.10a}
        \end{eqnarray}
in the Hilbert space with a simple $L^2$ inner product
        \begin{eqnarray}
        (\Psi_1|\Psi_2)_{\rm red}\equiv\int d\xi\,
        \Psi_1^*(\xi)\Psi_2(\xi).                    \label{5.13}
        \end{eqnarray}

The unitary map between the reduced phase space quantization of the above
type and the Dirac quantization of Sects. 2 -- 4  consists in a special
relation between the two-point kernel (\ref{4.3}) with prefactor (\ref{4.9})
and the Schrodinger evolution operator (\ref{5.10}) \cite{GenSem,BKr,BarvU}.
This relation is based on the equality of the principal Hamilton functions
in the original constrained theory and the reduced one and its corollary --
the relation between their Van-Vleck matrices
        \begin{eqnarray}
        &&S(t,\xi|t',\xi')={\mbox{\boldmath$S$}}(q,q')
        \Big|_{q=e(\xi,t),\,q'=e(\xi',t')},            \label{5.13a}\\
        &&{\mbox{\boldmath$S$}}_{ik'}e^i_A e^{k'}_{B'}=
        \frac{\partial^2 S(t,\xi|t',\xi')}
        {\partial\xi^A\,\partial\xi^{B'}}.
        \end{eqnarray}
Decomposing the gauge-fixed matrix (\ref{4.10}) in the basis of vectors
$(e^i_A,\nabla^i_\mu)$ and using (\ref{5.8}) one then easily finds the
needed relation \cite{GenSem,BKr,BarvU}
        \begin{eqnarray}
        K(t,\xi|t',\xi')={\rm const}\,
        \left.\left(\frac JM\right)^{1/2}
        {\mbox{\boldmath$K$}}(q,q')
        \left(\frac{J'}{M'}\right)^{1/2}
        \right|_{q=e(\xi,t),\,q'=e(\xi',t')}.          \label{5.9}
        \end{eqnarray}

This relation implies that the kernel ${\mbox{\boldmath$K$}}(q,q')$ similarly
to the Schrodinger propagator $K(t,\xi|t',\xi')$ can be regarded as a
propagator of the Dirac wavefunction ${\mbox{\boldmath$\Psi$}}(q)$ in
superspace. Indeed, introducing the following map between
${\mbox{\boldmath$\Psi$}}(q)$ and $\Psi(t,\xi)$
        \begin{eqnarray}
        \Psi(\xi,t)=\left.\left(\frac JM\right)^{1/2}\,
        {\mbox{\boldmath$\Psi$}}(q)\,
        \right|_{q=e(\xi,t)}                   \label{5.11}
        \end{eqnarray}
and taking into account the relation (\ref{5.5}) between the integration
measures on superspace and the physical space $\Sigma$, one finds that
the propagation law (\ref{5.10a}) in $\xi$-space can be regarded as a
projection onto $\Sigma(t)$ of the following propagation of the Dirac
wavefunction ${\mbox{\boldmath$\Psi$}}(q)$ from the initial Cauchy surface
$\Sigma(t')$ in $q$-space to the entire superspace
        \begin{eqnarray}
        {\mbox{\boldmath$\Psi$}}(q)=
        \int dq'\,{\mbox{\boldmath$K$}}(q,q')
        \,\delta(\chi(q',t'))
        J(q',-\partial{\mbox{\boldmath$S$}(q,q')}/\partial q')\,
        {\mbox{\boldmath$\Psi$}}(q')+O(\hbar).             \label{5.11a}
        \end{eqnarray}
Here the actual integration runs over the initial physical space $\Sigma(t')$.
However, the integration measure involves not just local quantities at
this surface, but also the normal derivatives of the kernel (or the
wavefunction ${\mbox{\boldmath$\Psi$}}(q')$ itself) arising in the one-loop
approximation as a Hamilton-Jacobi argument
$-\partial{\mbox{\boldmath$S$}(q,q')}/\partial q'$ of $J(q,p)$.

\subsection{General gauge conditions}
\hspace{\parindent}
Quantum reduction of the above type has been built in \cite{GenSem,BKr,BarvU}
for a particular case of coordinate gauge conditions, $\chi^\mu(q,t)=0$,
independent of momenta. The extension to general unitary gauges of the
form (\ref{5.03}) is less obvious, because such gauges no longer determine
the coordinate physical space, but rather a $(2n-m)$-dimensional subspace
in the entire pase space of $(q^i,p_i)$. Therefore, the additional reduction
of its dimensionality to $2(n-m)$ due to solving the constraints (\ref{1.2})
leaves us with the physical phase space without a natural decomposition into
coordinates and momenta. On the other hand, we need such a decomposition
(choice of polarization) for sake of quantization in coordinate
representation. This decomposition can, however, be built if we
model the $p$-dependent gauges by the coordinate ones and then repeate the
construction of the above type. The idea is that semiclassically, the
momenta equal the Hamilton-Jacobi values up to quantum fluctuations,
$p=\partial{\mbox{\boldmath$S$}}/\partial q+O(\hbar)$, so that the coordinate
gauges can be obtained from the general unitary gauges by the replacement
        \begin{eqnarray}
        \chi^\mu(q,p,t)\rightarrow\chi^\mu(q,t)=
        \chi^\mu\Big(q,\frac{\partial{\mbox{\boldmath$S$}}}
        {\partial q},t\Big).                            \label{5.11b}
        \end{eqnarray}
Then the reduction procedure is rather straightforward, although it involves
an additional step of the unitary transformation from the physical variables
in the initial gauge $\chi^\mu(q,p)=0$ to those of its coordinate version
(\ref{5.11b}).

The peculiarity of the reduction in $p$-dependent gauges is that, in
contrast with (\ref{5.8a}), the canonical transformation of the original
symplectic form $p_idq^i$ to the physical one $\pi_A d\xi^A$ is not contact.
It involves a nontrivial generating function ${\mbox{\boldmath$F$}}(q,\xi,t)$
        \begin{eqnarray}
        \Big(\,p_i dq^i-H_0dt\,\Big)_{\chi^\mu,\,T_\mu=0}=
        \pi_A d\xi^A-H_{\rm red}dt
        +d{\mbox{\boldmath$F$}}(q,\xi,t),                \label{5.24}
        \end{eqnarray}
so that the original action (\ref{1.1}) in terms of physical variables
acquires extra surface terms at the final and initial moments of time
$t_\pm$\footnote{For notational reasons in this section we supply the
arguments of two-point quantities by $\pm$ instead of primed and unprimed
labels above.}
        \begin{eqnarray}
        S[q,p]\Big|_{\chi^\mu,\,T_\mu=0}
        =\int_{t_-}^{t_+} dt\,(\pi_A\dot\xi^A-H_{\rm red})+
        {\mbox{\boldmath$F$}}(q_+,\xi_+,t_+)
        -{\mbox{\boldmath$F$}}(q_-,\xi_-,t_-).          \label{5.25}
        \end{eqnarray}
The actual form of the generating function ${\mbox{\boldmath$F$}}(q,\xi,t)$
depends on the choice of gauge conditions and the choice of physical
variables $(\xi^A,\pi_A)$. It can be rather complicated, but, fortunately,
we do not need it explicitly.

Important property of the surface terms in
(\ref{5.25}) is that they take care of the boundary conditions on the
histories $(q(t),p(t))$ at $t_\pm$. Natural boundary conditions for the
variational problem with the action (\ref{1.1}) -- left-hand side of
eq.(\ref{5.25}) -- correspond to fixed coordinates
        \begin{equation}
        q^i(t_\pm)=q^i_\pm                        \label{5.26}
        \end{equation}
and arbitrarily varied momenta. On the contrary, the variational problem for
the reduced phase-space action -- the integral term in the right-hand side
of (\ref{5.25}) -- assumes fixed physical coordinates
        \begin{equation}
        \xi^A(t_\pm)=\xi^A_\pm,                   \label{5.27}
        \end{equation}
but the reduction from $(q,p)$ to $(\xi,\pi)$ intertwines coordinates with
momenta, $q=q(\xi,\pi)$. Generating functions in the surface terms of
the right-hand side of (\ref{5.25}) reconcile these two different boundary
value problems.

Another important aspect of the boundary conditions (\ref{5.26}) is that
the values $q_\pm$ cannot be arbitrary. In complete analogy
with eq.(\ref{5.13a}) they should belong to the physical subspace of
the full phase space, that is to satisfy both the constraints and gauge
conditions. This property was obvious in coordinate gauges which directly
restricted possible values of $q_\pm$ to the surface (\ref{5.4}) of gauge
conditions. For $p$-dependent gauges one could think that the flexibility
of choosing the momenta at $t_\pm$ completely releases the values $q_\pm$
to be arbitrary. This is, however, not the case. One way to see this is
to calculate the action at the extremal satisfying classical equations of
motion and boundary conditions (\ref{5.26}). The resuling principal Hamilton
function ${\mbox{\boldmath$S$}}(q_+,q_-)$ uniquely determines the end-point
momenta, $p_\pm=\pm\partial{\mbox{\boldmath$S$}}/\partial q_\pm$, and the
end-point gauge conditions take the form of the {\em coordinate} gauges
of the form (\ref{5.11b})
        \begin{eqnarray}
        \chi^\mu_\pm(q_\pm)\equiv
        \chi^\mu(q_\pm,\pm\partial{\mbox{\boldmath$S$}}
        /\partial q_\pm,t_\pm)=0,                       \label{5.27a}
        \end{eqnarray}
that determine the embedding of the following two subspaces into the
{\em coordinate} space
        \begin{eqnarray}
        \Sigma_\pm: q^i_\pm=e^i(\eta^A_\pm,t_\pm),\,\,
        \chi^\mu_\pm(e^i(\eta,t_\pm))\equiv 0.            \label{5.28}
        \end{eqnarray}
Here the internal coordinates $\eta^A$ no longer coincide with the physical
coordinates $\xi^A$ of the reduced action.

Thus, the principal Hamilton function reduced to subspaces (\ref{5.28})
takes the form
        \begin{eqnarray}
        &&{\mbox{\boldmath$S$}}(q_+,q_-)
        \Big|_{q_\pm=e(\eta_\pm,t_\pm)}
        ={\cal S}(t_+,\eta_+|t_-,\eta_-),     \label{5.29}\\
        &&{\cal S}(t_+,\eta_+|t_-,\eta_-)
        =S(t_+,\xi_+|t_-,\xi_-)+F(\eta_+,\xi_+,t_+)
        -F(\eta_-,\xi_-,t_-),                          \label{5.30}
        \end{eqnarray}
where $S(t_+,\xi_+|t_-,\xi_-)$ is the principle Hamilton function of the
reduced system in the $p$-dependent gauge with fixed end-point coordinates
(\ref{5.27}) and
        \begin{eqnarray}
        F(\eta_\pm,\xi_\pm,t_\pm)=
        {\mbox{\boldmath$F$}}(e(\eta_\pm,t_\pm),\xi_\pm,t_\pm) \label{5.31}
        \end{eqnarray}
represent the generating functions of canonical transformations from
$\xi_\pm$ to $\eta_\pm$ at $t_\pm$. The whole expression on the right-hand
side is a function of $\eta_\pm$, the coordinates $\xi_\pm$ being expressed
in terms of $\eta_\pm$ as solution of equations
        \begin{eqnarray}
        \frac{\partial S(t_+,\xi_+|t_-,\xi_-)}{\partial\xi_\pm}\pm
        \frac{\partial F(\eta_\pm,\xi_\pm,t_\pm)}
        {\partial\xi_\pm}=0,                              \label{5.32}
        \end{eqnarray}
(these in turn follow from the corollary of eq.(\ref{5.24}),
$\pi_A=-\partial F(q,\xi,t)/\partial\xi^A$, and the Hamilton-Jacobi value
of the physical momentum,
$\pi_\pm=\pm\partial S(t_+,\xi_+|t_-,\xi_-) /\partial\xi_\pm$).

In this gauge we can now repeate the construction of the Dirac two-point
kernel of Sect.4. We identify the covectors $(\chi^\mu_i,\chi^\mu_{i'})$
with gradients of coordinate gauge conditions,
$\partial\chi^\mu_\pm(q)/\partial q^i$. Then we use (\ref{5.29}) and again
have the relation (\ref{5.9}) with the kernel $K(t,\xi|t',\xi')$ replaced by
        \begin{eqnarray}
        {\cal K}(t_+,\eta_+|t_-,\eta_-)\equiv
        \left[{\rm det}\,\frac i{2\pi\hbar}
        \frac{\partial^2{\cal S}(t_+,\eta_+|t_-,\eta_-)}
        {\partial\eta_+\;\partial\eta_-}\right]^{1/2}
        \,e^{\,{}^{\textstyle{\frac i{\hbar}
        {\cal S}(t_+,\eta_+|t_-,\eta_-)}}}              \label{5.33}
        \end{eqnarray}
and the factors $(J,J')=J_\pm$ (\ref{4.11}) calculated in the coordinate
gauges of the above type (\ref{5.27a}). Apriori, they differ from the
Faddeev-Popov determinants corresponding to $p$-dependent gauges, but as
shown in Appendix A, eq. (\ref{A7}),
        \begin{eqnarray}
        \left.\nabla^i_\nu\frac\partial{\partial q^i}
        \left[\,\chi^\mu\Big(q,\frac{\partial{\mbox{\boldmath$S$}}}
        {\partial q}\Big)\,\right]\equiv
        \Big\{\chi^\mu(q,p),T_\nu(q,p)\Big\}\,
        \right|_{\textstyle p=
        \partial{\mbox{\boldmath$S$}}/\partial q},           \label{5.23a}
        \end{eqnarray}
and all the $J^\mu_\nu$-factors coincide with true Faddeev-Popov matrices
in generic unitary gauges.

The interpretation of the kernel (\ref{5.33}) is obvious. It is built of the
principal Hamilton function (\ref{5.30}) corresponding to fixed boundary
values of $\eta$-variables at $t_\pm$. As discussed above, it contains
as total derivative terms the generating function
$F(\eta,\xi,t)$ of the canonical transformation from $\xi_\pm$ to $\eta_\pm$.
At the quantum level, in the semiclassical approximation linear in $\hbar$,
the kernel of the corresponding unitary transformation reads
        \begin{eqnarray}
        U(t,\eta|\xi)\equiv
        \left[{\rm det}\,\frac i{2\pi\hbar}
        \frac{\partial^2 F(\eta,\xi,t)}
        {\partial\eta\;\partial\xi}\right]^{1/2}
        \,e^{\,{}^{\textstyle{\frac i{\hbar}
        F(\eta,\xi,t)}}}.                               \label{5.34}
        \end{eqnarray}
If we represent the kernels of the above type as matrix elements of the
corresponding unitary operators in the coordinate ($\xi$ and $\eta$)
representations
        \begin{eqnarray}
        &&{\cal K}(t_+,\eta_+|t_-,\eta_-)
        =<\eta_+\,|\,\hat{\cal K}(t_+,t_-)\,|\,\eta_->, \\
        &&U(t,\eta|\xi)=<\eta\,|\,\hat U(t)\,|\,\xi>,   \\
        &&K(t_+,\xi_+|t_-,\xi_-)=<\xi_+\,|\,\hat K(t_+,t_-)\,|\,\xi_->,
        \end{eqnarray}
then the kernel in question (\ref{5.33}) reads as follows. It is nothing but
the unitary evolution operator of the reduced theory in generic $p$-dependent
gauge $\hat K(t_+,t_-)$ (whoes kernel is defined by eq.(5.10) with the
principal Hamilton function subject to fixed $(\xi,\xi')=\xi_\pm$) unitarily
transformed to the representation of $\eta$-variables
        \begin{eqnarray}
        \hat{\cal K}(t_+,t_-)=\hat U(t_+)\hat K(t_+,t_-)
        \hat U^\dagger(t_-).
        \end{eqnarray}
This unitary equivalence relation can be directly checked by the composition
of kernels (\ref{5.34}) and (\ref{5.10}) and calculating the corresponding
inegrals over $\xi_\pm$ by stationary phase technique\footnote
{The contribution of stationary points -- solutions of eqs.(\ref{5.32}) --
yields the prexponential factor of (\ref{5.33}), while its phase is achieved
as a linear combination of Hamilton-Jacobi phases (\ref{5.30}).}.

This reveals the role of extra unitary transformation arising
in the reduction to physical sector in generic momentum-dependent unitary
gauges (\ref{5.3}). A similar mechanism of modelling the non-unitary
(relativistic) gauges, which involve the Lagrange multiplyers and their
derivatives, by unitary ones was recently proposed in \cite{reduc} where
the Dirac two-point kernel was obtained by a direct calculation of the
one-loop path integral in gauge theory. The construction of such gauges is
based on the knowledge of solutions of linearized wave equations in a
corresponding relativistic gauge.

\section{The physical inner product: Hermiticity and gauge independence}
\hspace{\parindent}
The auxiliary inner product (\ref{2.8}) cannot serve as an inner product for
physical states because it is not well defined. In view of quantum constraints
the physical states have a distributional nature
$|\,{\mbox{\boldmath$\Psi$}}\big>="\delta(\hat{T})"
|\,{\mbox{\boldmath$\Psi$}}_{\rm aux}\big>$ with somehow determined
$m$-dimensional delta-function of non-abelian operators $\hat{T}_\mu$, and
their naive bilinear combinations are divergent because
$["\delta(\hat{T})"]^2\sim\delta(0)"\delta(\hat{T})"$. At most the auxiliary
vectors $|\,{\mbox{\boldmath$\Psi$}}_{\rm aux}\big>$ participating in the
construction of the physical states can be required to be square-integrable
in $L^2$ sense and thus induce a finite inner product for
$|\,{\mbox{\boldmath$\Psi$}}\big>$ (which is the idea of the so-called
refined algebraic quantization of constrained systems \cite{ALMMT,Land}).
Another approach consists in the unitary map from the Dirac to the reduced
phase space quantization of the previous section. Reduced theory has a
trivial inner product (\ref{5.13}) which induces a correct physical product in
the Dirac quantization scheme
        \begin{eqnarray}
        ({\mbox{\boldmath$\Psi$}}_2|{\mbox{\boldmath$\Psi$}}_1)
        =(\Psi_2|\Psi_1)_{\rm red}                        \label{5.12}
        \end{eqnarray}
on account of the relation (\ref{5.11}) and the
change of integration variables (\ref{5.6}). The result for semiclassical
states of the form (\ref{4.1}) looks like \cite{BPon,GenSem,BKr}
        \begin{eqnarray}
        ({\mbox{\boldmath$\Psi$}}_2|{\mbox{\boldmath$\Psi$}}_1)=
        \int dq\,
        {\mbox{\boldmath$\Psi$}}_2^*(q)\,\delta(\chi(q))
        J(q,\partial{\mbox{\boldmath$S$}}/\partial q)\,
        {\mbox{\boldmath$\Psi$}}_1(q)+O(\hbar).           \label{5.1}
        \end{eqnarray}
Here $\chi(q)=\chi^\mu(q)$ is a set of gauge conditions delta-function
of which
        \begin{eqnarray}
        \delta(\chi)=\prod_{\mu}\delta(\chi^\mu(q)),    \label{5.1a}
        \end{eqnarray}
determines the $(n-m)$-dimensional physical subspace $\Sigma$
embedded in superspace and $J(q,\partial{\mbox{\boldmath$S$}}/\partial q)$
is a corresponding Faddeev-Popov determinant\footnote
{In eq.(\ref{5.1}) the product
${\mbox{\boldmath$\Psi$}}_2^*{\mbox{\boldmath$\Psi$}}_1=
P_2^* P_1 \exp [i({\mbox{\boldmath$S$}}_1-{\mbox{\boldmath$S$}}_2)/\hbar]$
involves two different Hamilton-Jacobi functions, so that it seems ambiguous
on which Lagrangian manifold ($p=\partial{\mbox{\boldmath$S$}}_1/\partial q$
or $p=\partial{\mbox{\boldmath$S$}}_2/\partial q$) the relevant momentum
argument of $J(q,p)$ should be constructed. One should remember, however,
that in semiclassical expansion the integral is calculated by the
stationary phase method in which a dominant contribution comes from the
stationary point satisfying $\partial{\mbox{\boldmath$S$}}_1/\partial q=
\partial{\mbox{\boldmath$S$}}_2/\partial q$. This makes these Lagrangian
surfaces to coincide in the leading order, their difference being treated
perturbatively as expansion in $\hbar$.
}.

This explains the nature of the physical inner product in the Dirac
quantization. Its measure contains the Faddeev-Popov determinant which
depends on the choice of the constraint basis and transforms under the
transition (\ref{1.7}) to another basis as a density of weight 1 in
the space of gauge indices (\ref{4.12}) as compared to -1/2 weight of the
Dirac wavefunctions (\ref{3.8}). This proves the invariance of the
physical inner product under this transformation and shows that at the
quantum level it is not only canonical but also unitary. Our purpose now,
till the end of this section, will be to discuss the Hermiticity properties
of physical observables relative to this inner product and the gauge
independence of their matrix elements.

The semiclassical physical inner product (\ref{5.1}) can be rewritten as
an auxiliary inner product of physical states with a nontrivial
{\em operatorial} measure
        \begin{eqnarray}
        ({\mbox{\boldmath$\Psi$}}_1|
        {\mbox{\boldmath$\Psi$}}_2)=\big<{\mbox{\boldmath$\Psi$}}_1|
        \,\hat{J}\delta(\hat{\chi})\,
        |\,{\mbox{\boldmath$\Psi$}}_2\big>
        +O(\hbar).                                       \label{5.14}
        \end{eqnarray}
Here the operator ordering in operators of the ghost determinant and
gauge conditions is unimportant because it effects the multiloop orders
$O(\hbar)$ that go beyond the scope of this paper. The matrix element
of the physical observable $\hat{\cal O}_I$ is therefore
        \begin{eqnarray}
        ({\mbox{\boldmath$\Psi$}}_1|\,\hat{\cal O}_I\,|
        {\mbox{\boldmath$\Psi$}}_2)\equiv
        ({\mbox{\boldmath$\Psi$}}_1|\,\hat{\cal O}_I
        {\mbox{\boldmath$\Psi$}}_2)=
        \big<\,{\mbox{\boldmath$\Psi$}}_1|\,\hat{J}\,\delta(\hat{\chi})
        \hat{\cal O}_I
        |\,{\mbox{\boldmath$\Psi$}}_2\big>.              \label{5.15}
        \end{eqnarray}
To check the Hermiticity of $\hat{\cal O}_I$ we have to show that the
expression
        \begin{eqnarray}
        (\,\hat{\cal O}_I{\mbox{\boldmath$\Psi$}}_1|
        \,{\mbox{\boldmath$\Psi$}}_2)-
        ({\mbox{\boldmath$\Psi$}}_1|\,\hat{\cal O}_I
        {\mbox{\boldmath$\Psi$}}_2)=
        \big<\,{\mbox{\boldmath$\Psi$}}_1|\,
        \hat{\cal O}_I^{\dagger}\hat{J}\,\delta(\hat{\chi})
        |\,{\mbox{\boldmath$\Psi$}}_2\big>-
        \big<\,{\mbox{\boldmath$\Psi$}}_1|\,\hat{J}\,\delta(\hat{\chi})
        \hat{\cal O}_I
        |\,{\mbox{\boldmath$\Psi$}}_2\big>,               \label{5.16}
        \end{eqnarray}
where a dagger denotes Hermitian conjugation with respect to the auxiliary
inner product, is vanishing. From (\ref{2.12}) it follows that
$\hat{\cal O}_I^{\dagger}=\hat{\cal O}_I-i\hbar U^{\lambda}_{I\,\lambda}+
O(\hbar^2)$ (we consider the algebras of observables with $U^J_{IJ}=0$),
and the expression above takes the form
        \begin{eqnarray}
        \big<\,{\mbox{\boldmath$\Psi$}}_1|\,
        [\,\hat{\cal O}_I,\hat{J}\,\delta(\hat{\chi})\,]-
        i\hbar U^{\lambda}_{I\,\lambda}\,
        |\,{\mbox{\boldmath$\Psi$}}_2\big>=O(\hbar^2),   \label{5.17}
        \end{eqnarray}
which, as shown in Appendix A, is vanishing in the one-loop approximation.
Thus, the physical observables are semiclassically Hermitian with respect to
the physical inner product (\ref{5.1}).

Another important property of this product and matrix elements of observables
is their independence of the choice of gauge conditions $\chi^\mu(q)$
participating in their construction. The gauge independence of the inner
product itself is based, as shown in \cite{BKr,Op}, on the fact that it can be
rewritten as an integral over $(n-m)$-dimensional surface $\Sigma$ of certain
$(n-m)$-form which is closed in virtue of the Dirac constraints on physical
states
        \begin{eqnarray}
        &&({\mbox{\boldmath$\Psi$}}_2
        |{\mbox{\boldmath$\Psi$}}_1)=
        \int_{\Sigma} \omega^{(n-m)},\,\,\,\,
        d\omega^{(n-m)}=0.                       \label{5.19}
        \end{eqnarray}
It follows then from the Stokes theorem that this integral is
independent of the choice of $\Sigma$ or equivalently of the choice of gauge
conditions specifying the physical subspace. The form $\omega^{(n-m)}$ in
the one-loop approximation equals
        \begin{eqnarray}
        \omega^{(n-m)}=\frac{dq^{i_1}\wedge ...
        \wedge dq^{i_{n-m}}}{(n-m)!} \epsilon_{i_1 ...i_n}
        {\mbox{\boldmath$\Psi$}}_2^*\,\nabla^{i_{n-m+1}}_{1} ...
        \nabla^{i_{n}}_{m}\,{\mbox{\boldmath$\Psi$}}_1,   \label{5.20}                    \label{3.59}
        \end{eqnarray}
and its closure is a corollary of the continuity equation (\ref{4.5})
for ${\mbox{\boldmath$\Psi$}}_2^*{\mbox{\boldmath$\Psi$}}_1$
        \begin{eqnarray}
        \frac\partial{\partial q^i}
        (\nabla^i_\mu {\mbox{\boldmath$\Psi$}}_2^*
        {\mbox{\boldmath$\Psi$}}_1)=
        U^{\lambda}_{\mu\lambda}
        {\mbox{\boldmath$\Psi$}}_2^*
        {\mbox{\boldmath$\Psi$}}_1.                      \label{5.21}
        \end{eqnarray}
The evaluation of the matrix element of the observable $\hat{\cal O}$
semiclassically involves the evaluation of the quantity
        \begin{eqnarray}
        ({\mbox{\boldmath$\Psi$}}_2
        |\,\hat{\cal O}_I{\mbox{\boldmath$\Psi$}}_1)=
        \int_{\Sigma} \omega^{(n-m)}
        {\cal O}_I(q,\partial{\mbox{\boldmath$S$}}_1/\partial q)+
        O(\hbar),                                        \label{5.22}
        \end{eqnarray}
which will be also gauge independent provided the continuity equation holds
for the quantity ${\mbox{\boldmath$\Psi$}}_2^*(q)
{\mbox{\boldmath$\Psi$}}_1(q)\,
{\cal O}_I(q,\partial{\mbox{\boldmath$S$}}_1/\partial q)$. But this equation
will also be a corollary of (\ref{5.21}) because, as shown in Appendix A,
on the Lagrangian manifold of phase space
        \begin{eqnarray}
        \left.\nabla^i_\mu\frac\partial{\partial q^i}
        \left[\,{\cal O}\left(q,\frac{\partial{\mbox{\boldmath$S$}}}
        {\partial q}\right)\,\right]\equiv\Big\{{\cal O},T_\mu\Big\}\,
        \right|_{\textstyle p=
        \partial{\mbox{\boldmath$S$}}/\partial q}=0           \label{5.23}
        \end{eqnarray}
in view of gauge invariance (\ref{2.10}) of observables\footnote
{In eqs.(\ref{5.20})-(\ref{5.22}) the difference between the momentum
arguments related to different Hamilton-Jacobi functions
$p=\partial{\mbox{\boldmath$S$}}_1/\partial q$
and $p=\partial{\mbox{\boldmath$S$}}_2/\partial q$ should also be treated
perturbatively in $\hbar$ and, thus, goes beyond the one-loop approximation
(see the footnote to eqs.(\ref{5.1})-(\ref{5.1a})).
}.
Thus, as it should have been expected from the theory of gauge fields
\cite{DW,FP} the gauge independence of the physical matrix elements or
expectation values of observables follows from the gauge invariance of
the latter. This is true not only at the formal path-integral quantization
level, but also in the operatorial Dirac quantization scheme.

\section{Conclusions}
\hspace{\parindent}
Thus we see that, despite a complicated non-abelian nature of the formalism,
Dirac quantization of generic constrained systems is remarkably consistent
and reveals rich geometrical structures beyond the lowest order semiclassical
approximation. Geometrical covariance of operators and physical states
takes place not only in the coordinate configuration space of the theory,
but also in the space of gauge transformations. The quantum formalism turns
out to be covariant in the sense of unitary equivalence with respect to
generic symmetries of a classical theory including arbitrary change of the
constraint basis. The operators of physical observables turn out to be
Hermitian with respect to a physical inner product and their matrix elements
are gauge independent in full correspondence with similar properties in
the classical domain. All these properties were obtained perturbatively in
the one-loop approximation of semiclassical expansion but, no doubt, they
can be extended to multi-loop orders, though, apparently by the price of
growing technical complexity.

It should be emphasized that this remarkably general picture of quantum
invariances was obtained in the {\em Dirac} quantization of constrained
systems, when {\em all} quantum constraints are imposed on physical
states. These states can be regarded a quantum truncation of the BFV(BRST)
quantization with ${\cal C}\bar{\cal P}$-ordered form of the nilpotent BRS
operator in the extended relativistic phase space of original $(q,p)$ and
ghost canonical variables $({\cal C},\bar{\cal P})$ \cite{BarvU,BKr,BatMarn}.
As mentioned in Introduction, the main motivation for such a setting is the
quantization of zero modes of extended objects for which another (unitary
inequivalent) quantization in the Fock space based on the normal Wick
ordering in the ghost sector is not applicable. The latter is usually
applied to oscillatory modes of field theoretical models, including
strings and low-dimensional CFT \cite{Brink}. This quantization after
truncation to $(q,p)$-sector results only in a half of the initial
constraints imposed on physical states \cite{BFV} (and in view of this
admitting the central extension of quantum algebra).  For them the
geometric properties of the above type are much less known if at all
available at such a general level not resorting to a concrete harmonic
oscillator decomposition of fields. Thus it seems interesting to try
extending the above geometric methods to this quantization scheme, although
it is very hard to expect that the general coordinate invariance in curved
configuration space of the theory can be fruitfully combined with the
Fock space representation. (The latter, as is well known, is basically
applicable to linearized fields belonging to the tangent vector space rather
than to the configuration space manifold.)

Another difficulty with the field theoretical (infinite dimensional)
extension of the above technique consists in ultraviolet infinities and the
accompanying them anomalies that would generally violate gauge symmetries
at the quantum level. As is well known, this problem cannot be analyzed at
the level of generality adopted in this paper, that is for gauge theories
of general type. A usual approach would consist in applying the above
formalism to a given field model within a particular regularization scheme
that would render all the quantities finite, singling out the potential
anomalies of gauge symmetries and using the results as selection criteria for
viable systems. Like in string models this might lead to restrictions on
the number of fields (dimensionality of target spacetime) or other parameters
of quantized model. More detailed discussion of this issue can be found in
\cite{BKr,Op}.

As far as it concerns the present results, their direct implications can
be expected in quantization of zero modes in various field theoretical
models or extended objects. In quantum cosmological context the Dirac
quantization scheme is implemented in the system of Wheeler-DeWitt
equations. The minisuperspace (zero) mode of the full superspace of
quantum cosmology cannot be quantized in the Fock representation, because
it has a ghost nature \cite{RubSved,GMTS,Turok}, so only the coordinate
representation and relevant methods of this paper can be used for the
studies of the quantum Cauchy problem for the early inflationary Universe
\cite{BarvU,qcr}. The gauge invariant observables in early cosmology, the
theory of which is very important for understanding the formation of
structure \cite{MukhFB} and back reaction phenomena, should be analyzed
by the technique of the above type. In this respect, especially important
becomes the reduction technique in generic unitary gauges that should be
applied in the theory of cosmological perturbations of \cite{GMTS} in
order to study the quantum back reaction effects in effective equations of
inflationary dynamics \cite{progress}. The potential range of applications
is even wider in D-brane dynamics of string theory \cite{Polch} and brane
cosmology \cite{brane}. Zero modes of D and p-branes in nonperturbative
string theory or brane-worlds in cosmology \cite{GT}, separation of their
dynamical from purely gauge properties, description of brane dynamics in
the bulk in the formalism of quantum Dirac constraints and many other issues
are subject to the methods of this paper.

\appendix
\renewcommand{\thesection}{\Alph{section}.}
\renewcommand{\theequation}{\Alph{section}.\arabic{equation}}
\section{Hermiticity of observables}
\hspace{\parindent}
To prove eq.(\ref{5.17}) note that
        \begin{eqnarray}
        \big<\,{\mbox{\boldmath$\Psi$}}_1|\,
        [\,\hat{\cal O}_I,\hat{J}\,\delta(\hat{\chi})\,]
        |\,{\mbox{\boldmath$\Psi$}}_2\big>=
        i\hbar\int dq\,\{\,{\cal O}_I,J\delta(\chi)\}\,
        {\mbox{\boldmath$\Psi$}}_1^*{\mbox{\boldmath$\Psi$}}_2+
        O(\hbar^2).                       \label{A1}
        \end{eqnarray}
The Poisson bracket commutator here can be transformed by using the cyclic
Jacobi identity and the weak gauge invariance of constraints (\ref{2.10})
        \begin{eqnarray}
        &&\int dq\,\{\,{\cal O}_I,J\delta(\chi)\}\,
        {\mbox{\boldmath$\Psi$}}_1^*{\mbox{\boldmath$\Psi$}}_2=
        \int dq\,\delta(\chi) J\,\left[\,U^\mu_{I\,\mu}+
        J^{-1\,\mu}_{\,\,\,\,\,\,\nu}\{T_\mu,\{\chi^\nu,{\cal O}_I\}\}\,
        \right]\,{\mbox{\boldmath$\Psi$}}_1^*{\mbox{\boldmath$\Psi$}}_2
          \nonumber\\
        &&\qquad\qquad\qquad\qquad\qquad
        +\int dq\,\frac{\partial\delta(\chi)}{\partial\chi^\mu} J\,
        \{{\cal O}_I,\chi^\mu\}\,
        {\mbox{\boldmath$\Psi$}}_1^*
        {\mbox{\boldmath$\Psi$}}_2.              \label{A2}
        \end{eqnarray}
The last term here can be integrated in a special coordinate system on
superspace defined by eqs.(\ref{5.3}) and (\ref{5.5})
        \begin{eqnarray}
        \int dq\,\frac{\partial\delta(\chi)}{\partial\chi^\mu} J\,
        \{{\cal O}_I,\chi^\mu\}\,
        {\mbox{\boldmath$\Psi$}}_1^*{\mbox{\boldmath$\Psi$}}_2=
        -\int dq\,\delta(\chi)\,M\,\frac{\partial}{\partial\theta^\mu}
        \left[\,\frac JM\,\{{\cal O}_I,\chi^\mu\}\,
        {\mbox{\boldmath$\Psi$}}_1^*
        {\mbox{\boldmath$\Psi$}}_2\,\right].              \label{A3}
        \end{eqnarray}

To transform this expression further let us derive several useful identities.
First of all, note that any function of phase space variables $f(q,p)$ when
restricted to the Lagrangian manifold defined by the Hamilton-Jacobi
function $\mbox{\boldmath$S$}$ becomes a function on superspace
$f(q,\partial{\mbox{\boldmath$S$}}/\partial q)$. Its derivative with
respect to coordinate $\theta^\mu$ (of the new coordinate system)
        \begin{eqnarray}
        \frac{\partial}{\partial\theta^\mu}
        \left[\,f\left(q,\frac{\partial{\mbox{\boldmath$S$}}}{\partial q}
        \right)\,\right]=e^i_\mu\left(\frac{\partial f}{\partial q^i}+
        \frac{\partial f}{\partial p_k}
        \frac{\partial^2\mbox{\boldmath$S$}}
        {\partial q^k \,\partial q^i}\right),            \label{A4}
        \end{eqnarray}
can be simplified to
        \begin{eqnarray}
        \frac{\partial}{\partial\theta^\mu}
        \left[\,f\left(q,\frac{\partial{\mbox{\boldmath$S$}}}{\partial q}
        \right)\,\right]=J^{-1\,\nu}_{\,\,\,\,\,\,\mu}\{f,T_\nu\}+
        \Omega^A_\mu\frac{\partial}{\partial\xi^A}
        \left[\,f\left(q,\frac{\partial{\mbox{\boldmath$S$}}}
        {\partial q}\right)\,\right].                     \label{A6}
        \end{eqnarray}
in view of eq.(\ref{5.7}) and the differentiated version of the
Hamilton-Jacobi form of constraints
        \begin{eqnarray}
        \nabla^i_\mu \frac{\partial^2\mbox{\boldmath$S$}}
        {\partial q^i \,\partial q^k}=
        -\frac{\partial T_\mu}{\partial q^k}.         \label{A5}
        \end{eqnarray}
Similar derivation shows that the gauge transformation of this function
on the Lagrangian manifold, generated by the vector flow $\nabla^i_\mu$,
coincides with the Poisson bracket of $f(q,p)$ with the constraint
        \begin{eqnarray}
        \nabla^i_\mu \frac{\partial}{\partial q^i}
        \left[\,f\left(q,\frac{\partial{\mbox{\boldmath$S$}}}
        {\partial q}
        \right)\,\right]=\{f,T_\mu\},                  \label{A7}
        \end{eqnarray}
evaluated certainly at $p=\partial{\mbox{\boldmath$S$}}/\partial q$. With
these identities and using the derivatives of the measure $M$ (\ref{5.5})
        \begin{eqnarray}
        \frac {\partial M}{\partial \theta^\mu}=
        -M\frac{\partial e^i_\mu}{\partial q^i},\,\,\,\,
        \frac {\partial M}{\partial \xi^A}=
        -M\frac{\partial e^i_A}{\partial q^i}        \label{A8}
        \end{eqnarray}
($e^i_\mu$ and $e^i_A$ are defined by (\ref{5.5a})), the relation (\ref{5.7})
and the Jacobi identity for Poisson brackets one can obtain the following
gauge derivative
        \begin{eqnarray}
        \frac {\partial}{\partial \theta^\mu}\left(\frac JM \right)
        =\frac JM\,J^{-1\,\nu}_{\,\,\,\,\,\,\mu}\left(
        \frac{\partial\nabla^i_\nu}{\partial q^i}
        -U^\lambda_{\nu\lambda}\right)+
        \frac {\partial}{\partial \xi^A}
        \left(\frac JM \Omega^A_\mu\right).         \label{A9}
        \end{eqnarray}
Using this equation for $J/M$ and applying (\ref{A5}) and (\ref{A7})
to other quantities in (\ref{A3}) we finally obtain
        \begin{eqnarray}
        &&\int dq\,\delta(\chi)\,M\,\frac{\partial}{\partial\theta^\mu}
        \left[\,\frac JM\,\{{\cal O}_I,\chi^\mu\}\,
        {\mbox{\boldmath$\Psi$}}_1^*
        {\mbox{\boldmath$\Psi$}}_2\,\right]=
        \int dq\,\delta(\chi)\, JJ^{-1\,\mu}_{\,\,\,\,\,\,\nu}
        \{T_\mu,\{\chi^\nu,{\cal O}_I\}\}\,
        {\mbox{\boldmath$\Psi$}}_1^*{\mbox{\boldmath$\Psi$}}_2
        \nonumber\\
        &&\qquad\qquad\qquad\qquad+\int dq\,\delta(\chi)\, J
        J^{-1\,\mu}_{\,\,\,\,\,\,\nu}\{{\cal O}_I,\chi^\nu\}\,\left[
        \frac \partial{\partial q^i}
        (\nabla^i_\mu {\mbox{\boldmath$\Psi$}}_1^*
        {\mbox{\boldmath$\Psi$}}_2)
        -U^\lambda_{\mu\lambda}
        {\mbox{\boldmath$\Psi$}}_1^*{\mbox{\boldmath$\Psi$}}_2\right]
        \nonumber\\
        &&\qquad\qquad\qquad\qquad
        +\int_{\Sigma} d\xi\,\frac {\partial}{\partial \xi^A}
        \left[\,\frac JM\Omega^A_\mu\{{\cal O}_I,\chi^\mu\}
        {\mbox{\boldmath$\Psi$}}_1^*
        {\mbox{\boldmath$\Psi$}}_2\,\right].            \label{A10}
        \end{eqnarray}
The second term here is vanishing in view of the continuity equation
(or more precisely $O(\hbar)$ for different ${\mbox{\boldmath$\Psi$}}_1$
and ${\mbox{\boldmath$\Psi$}}_2$). The third total derivative term is
vanishing in view of zero boundary conditions for physical wavefunctions
at the infinity of the physical space $\Sigma$. Collecting equations
(\ref{A2}), (\ref{A3}) and (\ref{A10}) together we see that the matrix
element of the commutator reduces to the one term containing the trace of the
structure function $U^\mu_{I\,\mu}$ which exactly cancels out in the
equation (\ref{5.17}). This proves the Hermiticity of observables in the
physical inner product.

\section{Abelianization procedure on Lagrangian manifolds}
\hspace{\parindent}
It is well known that when the structure functions of the Poisson-bracket
algebra of constraints (\ref{1.3}) are not constants this algebra is open
\cite{BV}: the commutator of two consequitive transformations
$\delta_\mu f\equiv\{f,T_\mu\}$ of any function on phase space
        \begin{eqnarray}
        \{\{f,T_\mu\},T_\nu\}-\{\{f,T_\nu\},T_\mu\}=
        U^\lambda_{\mu\nu}\,\{f,T_\lambda\}
        +\{f,U^\lambda_{\mu\nu}\}\,T_\lambda      \label{B1}
        \end{eqnarray}
is a linear combination of these transformations only on the constraint
surface $T_\lambda(q,p)=0$. Semiclassically, the restriction to this surface
takes place on the Lagrangian manifold of phase space
        \begin{eqnarray}
        p_i=\frac{\partial{\mbox{\boldmath$S$}}}
        {\partial q^i}                                   \label{B2}
        \end{eqnarray}
defined by the Hamilton-Jacobi function of a semiclassical state, satisfying
the Hamilton-Jacobi equation (\ref{4.4}). As was shown in Appendix A, the
action of the constraint generators on $f(q,p)$ (in the sense of Poisson
brackets) on this surface (\ref{A7}) can be generated by directional
derivatives along a special set of vectors $\nabla^i_\mu$ (\ref{4.6}). Therefore
these vectors can be regarded as gauge generators on the Lagrangian manifold
of phase space. They have a closed Lie-bracket algebra \cite{BKr}
        \begin{eqnarray}
        \nabla^i_\mu\frac{\partial\nabla^k_\nu}{\partial q^i}-
        \nabla^i_\nu\frac{\partial\nabla^k_\mu}{\partial q^i}=
        U^\lambda_{\mu\nu}\nabla^k_\lambda               \label{B3}
        \end{eqnarray}
and, therefore, can be abelianized \cite{Abelianization} by recombining these
vectors with the aid of the matrix of the inverse Faddeev-Popov operator
constructed out of some admissible gauge conditions $J^\mu_\nu\equiv
\nabla^i_\nu \partial\chi^\mu/\partial q^i$
        \begin{eqnarray}
        {\cal R}^i_\mu=J^{-1\,\nu}_{\,\,\,\,\,\,\mu}\nabla^i_\mu,\,\,\,\,
        \left[\,{\cal R}^i_\mu\frac{\partial}{\partial q^i},
        {\cal R}^i_\nu\frac{\partial}{\partial q^i}\,\right]=0.   \label{B4}
        \end{eqnarray}

Abelian generators alow one to construct preferred parametrization of
the coordinate manifold (\ref{5.3}) with a special internal coordinates on
a physical space $\Sigma$. Note that the new coordinates $\xi^A$ in
(\ref{5.3}) as functions of the original coordinates are not necessarily
gauge invariant. Abelianization procedure can render them invariant as
follows. Demand that the reparametrization functions in (\ref{5.3})
satisfy the equations
        \begin{eqnarray}
        \frac{\partial e^i(\xi,\theta^\mu)}{\partial\theta^\mu}=
        {\cal R}^i_\mu,                               \label{B5}
        \end{eqnarray}
which are integrable in view of the abelian nature of ${\cal R}^i_\mu$.
Then the identity $e^i_\mu \partial\xi^A/\partial q^i=0$ implies
the gauge invariance of $\xi^A(q)$
        \begin{eqnarray}
        \nabla^i_\mu\frac{\partial\xi^A}{\partial q^i}=0.    \label{B6}
        \end{eqnarray}
Many equations above simplify with this preferred parametrization of
physical space $\Sigma$. Indeed, eq.(\ref{B5}) implies that $\Omega^A_\mu=0$
in the equation (\ref{5.7}) and the equation (\ref{A9}) for the factor
$J/M$ performing
the map from the Dirac quantization to the reduced phase space quantization
takes the form of the continutity equation for a one-loop prefactor
        \begin{eqnarray}
        \frac {\partial}{\partial q^i}
        \left(\,\nabla^i_\mu \, J/M\,\right)=
        U^\lambda_{\mu\lambda}
        \,J/M,                      \label{B7}
        \end{eqnarray}
which means that this factor with the measure $M$ constructed in this
parametrization represents a {\em kinematical} solution of the continuity
equation.

\section*{Acknowledgements}
\hspace{\parindent}
The author is grateful to the participants of the workshop ``Mathematical
Problems in Quantum Gravity'' for helpful stimulating discussions and for a
financial support by the Erwin Schrodinger International Institute for
Mathematical Physics where an early stage of this work has been accomplished.
This work was also supported by the Russian Foundation for Basic Research
under grants 99-02-16122 and the European Community Grant INTAS-93-493-ext.
Partly this work has been made possible also due to the support by the
Russian Research Project ``Cosmomicrophysics''and the grant for support of
leading scientific schools 00-15-96699.

\end{document}